# Possibility and prevention of inappropriate data manipulation in Polar Data Journal

Takeshi Terui
National Institute of Polar Research
Research Organization of Information and Systems
Tachikawa, Japan
0000-0002-0231-2193

Yasuyuki Minamiyama
Policy Data Lab
The Tokyo Foundation for Policy Research
Minato-ku, Japan
0000-0002-7280-3342

Kazutsuna Yamaji
National Institute of Informatics
Research Organization of Information and Systems
Chiyoda-ku, Japan
0000-0001-6108-9385

*Abstract*—Stakeholders in the scientific field must always maintain transparency in the process of publishing research results in journals. Unfortunately, although research misconduct has stopped, certain forms of manipulation continue to appear in other forms. As new techniques of scientific publishing develop, science stakeholders need to examine the possibility of inappropriate activity in these new platforms. The National Institute of Polar Research in Japan launched a new data journal Polar Data Journal (PDJ) in 2017 to review the quality of data obtained in the polar region. To maintain transparency in this new data journal, we investigated the possibility of inappropriate data manipulation in peer reviews before the inception of this journal. We clarified inappropriate activity for the data in the peer review and considered preventive measures. We designed a specific workflow for PDJ. This included two measures: (i) the comparison of hash values in the review process and (ii) open peer review report publishing. Using the hash value comparison, we detected two instances of inappropriate data manipulation after the start of the journal. This research will help improve workflow in data journals and data repositories.

## I. Introduction

All submitted data is reviewed for quality and provenance checks during the review process of the data journal. Digital data would have a bigger size or a higher resolution than the printed format in regular scientific papers. Digital data can be easily changed, moved, or copied on the personal computer, even when this data pertains to big data. Therefore, data journal publishers must pay special attention to inappropriate data manipulation done during peer reviews and post-publishing reviews. Earlier, it was assumed that improper operations were mostly done on figures and tables in regular papers. Therefore, the checking system depended on human readability and the discrimination ability of the peer reviewers. However, considered it impossible to check the whole data in a data journal by using the review process followed for regular papers. Therefore, it became necessary to have a data journal-specific review process and rules.

Polar Data Journal (PDJ) is a data journal launched by the National Institute of Polar Research in Japan in January 2017 [1]. PDJ performs the quality control of various data observed in the polar regions using peer review; therefore, the stakeholders including scientists can confidently use the published data. When we developed the review process in PDJ, we listed the possibility of where the inappropriate data manipulation might occur in each submission process until the publication, and we considered the measures necessary for the listed risk.

By developing the framework of data publishing as per the recommendations of the Research Data Alliance standards, we designed a specific review process for PDJ to prevent inappropriate data manipulation [2]. The first step in the process was to calculate the hash value to confirm the identity of the data. The second step was to publish a peer review report on the PDJ web site. PDJ is an advanced data journal, which implements these processes for the official peer review process.

In this research, we investigate the various kinds of inappropriate data manipulations and examine how to prevent these manipulations. We have succeeded in detecting data manipulation twice after the launch of PDJ. We show two case studies about prevention and discuss the influence of the new method in data journals.

## II. Method

Fig. 1 is the latest flowchart about PDJ from the submission of a paper until the publication. PDJ uses the Editorial Manager (EM in Fig. 1) for the peer review process and uses the JAIRO Cloud [3] system for paper publications. PDJ does not have the original data repository service and uses external data repository services. This figure is included in the PDJ policy and must be followed as a workflow. PDJ also has an authorship policy and a data policy.

### A. Hash Value

The hash value is a typical value that is output when specific data is input into the hash function. This value is unique according to the input data; therefore, it is possible to detect tampering by comparing the hash values. PDJ uses SHA-256 as a hash function, which has a certification from CRYPTREC in Japan [4]. The input value of the hash function is the target data of the submitted manuscript. After copying this data from the data repository to the JAIRO Cloud, the hash values of those data are compared by using the PDJ secretariat (sea data flow (4), (4'), and (14) in Fig. 1). This comparison process makes it possible to know whether the data has been changed in each review state.

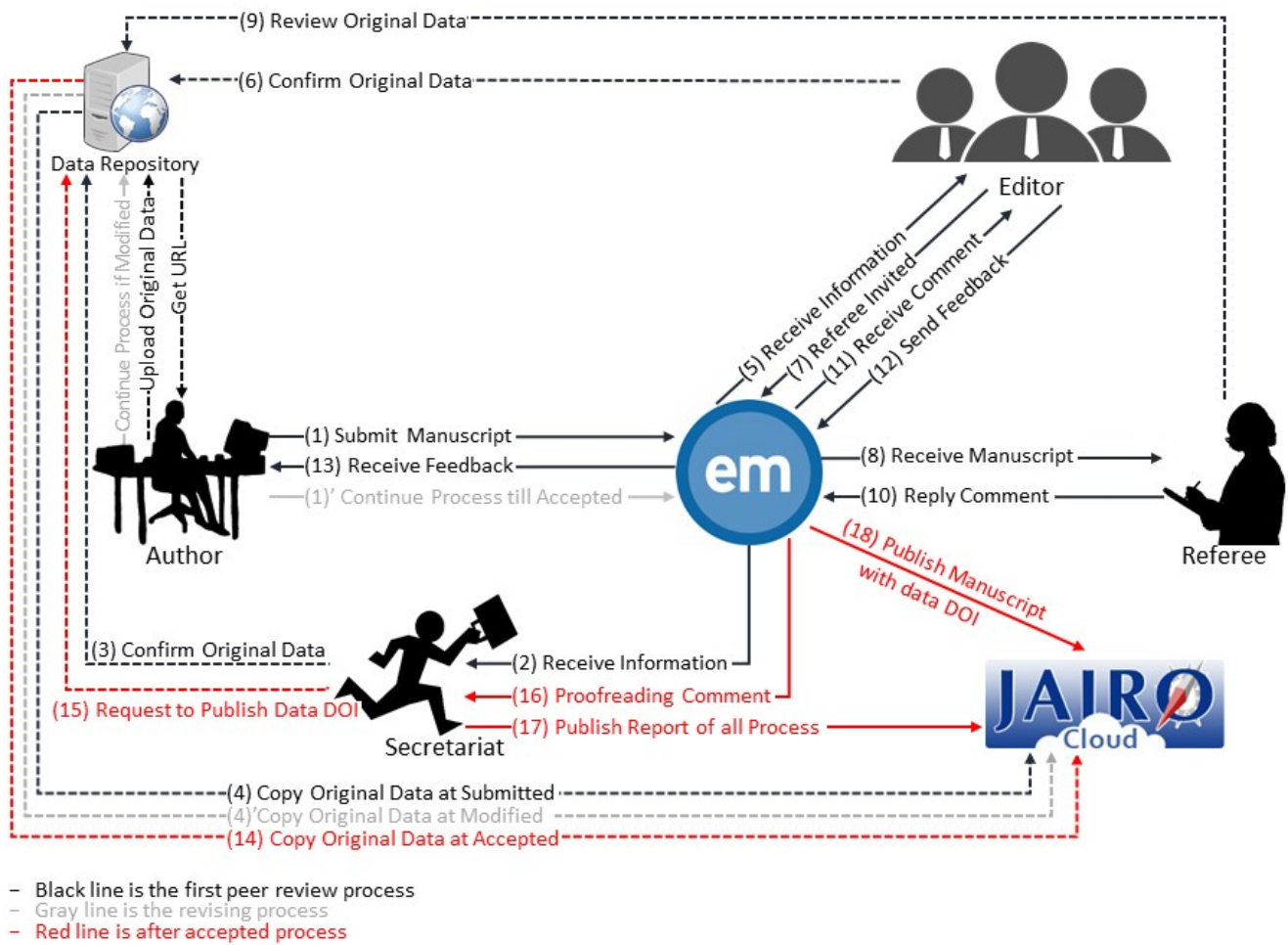

Fig. 1. The peer review process diagram of Polar Data Journal updated from reference [1]

*B. Peer Review Report*

The peer review report is a document that includes all comments and feedback in the review process. The open peer review report is known as PeerJ, it is the journal of life and environmental sciences [5]. Nature Communications started to provide the option of publishing the reviewer reports in 2015 [6]. The open peer review report is expected as a new approach to maintain transparency. In PDJ, the peer review report was published when the author's paper was published. The name and affiliation of the referee are disclosed in the report with the approval of the referee. It also includes the hash values and the download links for the data.

### III. INAPPROPRIATE DATA MANIPULATION

Inappropriate data manipulation is the fabrication, falsification, and plagiarism (FFP) of data. Digital data can be easily edited by using a personal computer. Inappropriate data manipulation may be done by stakeholders of the manuscript. The PDJ editorial committee will judge whether the act of manipulation was intentional or unintentional after a detailed investigation. However, we must develop a method to detect the FFP in the system. Therefore, we need to know about the situation in which inappropriate data manipulation might occur during the review process. We considered the possibility of data manipulation in each role in Fig. 1; this information has been summarized in Table 1. Based on this result, we considered how the hash value and the peer review report are useful for detecting each possibility.

*A. Author*

The author is the owner of both the submitted manuscript in the journal and the registered data in the data repository. The author has the authority for data manipulation and operation. The author must describe how to create data and how to use it. Before submitting a manuscript, the data must be registered with the data repository. The manuscript with an accessible

URL of the data is sent into the manuscript submission system by the author (Flow 1 in Fig. 1). The author is the owner of data; therefore, the following operations are available:

- Fake data registration.
- Unauthorized data change during the review process.
- Data change after the acceptance of the paper.

The judgment of whether data is fake or not is fundamentally a check made by the referee; this process is similar to the process followed for ordinary journals. The hash value comparison does not work for fake data registration because there is no comparable hash value. We considered that the strange data including the false data would be screened out by the referee's comments. The PDJ policy is defined so that the data format is the standard format used by the scientific community; if not, the data usage must be described. Although this policy is for quality improvement and data reusability, the amount of work for the referee may be reduced by this policy.

The referee and the editor do not allow any modifications during the review period without authorization in the standard journal. Changing the data without the consent of the referee or the editor should not also be done in the data journal. However, the possibility of data manipulation due to the author's mistakes cannot be excluded; therefore, data manipulation must be detected in the review process. We designed our process to confirm the identity of the data at the time of submission, revision, and acceptance of data (see data flow (4), (4'), and (14) in Fig. 1) by copying the data from the data repository and confirming the hash value.

Data renewal or update may occur after the manuscript acceptance due to the progress in research. If the author changes the data on the data repository side, it means that un-reviewed data had been published; this was not consistent with the paper. Therefore, we need a solution for end users to detect whether or not the data is peer-reviewed. PDJ's peer review report includes a permanent download link and the hash value of the reviewed data. End users can know the hash value from the report, and they can confirm whether or not the most recent data in the data repository is the same.

### B. Referee

The referee has the responsibility for checking the content of the manuscript and the data quality. PDJ have adopted a single-blind peer review process. The referee can perform the following operations on the data:

- Data plagiarism.
- Comments that induce data edits for the referee's benefit.

Unfortunately, peer reviewers have been plagiarizing [7]. Data copying is easy because a replica can be created by just a drag-and-drop operation. When the referee confirms the data, the referee performs this operation (see data flow (8) and (9) in Fig. 1). Therefore, it is possible to progress the referee's research by using this data because the referee has access to the earlier data also. This act may be suspected to be data plagiarism. If the data is unpublished, it is just plagiarism in appearance even if the referee does not deem it to be so. To overcome this, the data repository requirement needs various features (e.g., previewing the Web browser, generating temporary download links, and tracking data). However, not all data repositories worldwide have these functions. It is very unreasonable to demand complete confidentiality of the data seen by a third party. We concluded that it is difficult to detect such referee-side actions in the review process because they are carried out in the referee's environment. There was no technical solution; therefore, we considered ways to reduce plagiarism using policy restrictions. We designed the PDJ rule to require open access for the reviewing data. The referee will lose the advantages and motivations of early access by using open access.

The referee can indirectly induce data editing in the comments to the author (see data flow (10) in Fig. 1). Any comments that improve the data quality is useful. However, comments that benefit the referee are not beneficial, such as a comment that can induce a specific data format conversion for the referee's computer environment. The editor should prevent inductive comments that only benefit the referee, but it is not easy to find it in the specific scientific data because of its high expertise. We designed the peer-review report because suppressing such non-reasonable comments can be expected by including all the referee comments.

TABLE I. EACH ROLE, FUNCTION, INAPPROPRIATE DATA MANIPULATION, THE PROCESS OCCURRED IN FIG. 1, AND MEASURES IN PDJ

| Role | Functions | Inappropriate Data Manipulation | Flow in Fig. 1 | Measures in PDJ |
|---|---|---|---|---|
| Author | Registering data into a trust data repository Submitting a manuscript with data URL Revising the manuscript and data | Fake data registration | Pre-1 | Journal policy |
| | | Unauthorized data change during the review process | 2-13 | Hash value |
| | | Data change after the acceptance of the paper | 14-17 | Hash value |
| Referee | Reviewing a submitted paper and its data | Data plagiarism | 9 | Journal policy |
| | | Comments that induce data edits for the referee's benefit | 10 | Peer review report |
| Editor | Nominating the referee Judging manuscript accepted or not | Inappropriate referee nomination | 7 | Peer review report |
| | | Notification of inappropriate review results | 12 | Peer review report |
| Data Repository | Publishing data Archiving data Providing the landing page | Data loss | Mainly after 13 | Hash value |
| | | Data falsification | Mainly after 13 | Hash value |
| | | Data fabrication | Pre-1 and after 13 | Journal policy |
| Secretariat | Supporting the peer review process | Procedural error | 2-17 | System implementation |

*C. Editor*

The editor plays the most responsible role among the journal stakeholders. Editors have substantial authority over both the author and the peer reviewer. Editors cannot directly manipulate data, but they can use their authority to contribute to inappropriate data manipulations as follows:

- Inappropriate referee nomination.
- Notification of inappropriate review results.

Inappropriate referee nomination (see data flow 7 in Fig. 1) will result in inappropriate data manipulation by the referee (see Section III-B). It is also difficult to detect this activity in the review process. The peer review report always describes the editor's name and feedback. If the editor suspected any manipulation, it is possible to trace it back from the peer-review report.

The editor has strong authority over the selection of peer review results (see data flow (12) in Fig 1). If the editor reports a review result that has lost neutrality, there is no way to prevent the FFP. Both mutual oversight and strong governance by the editorial board would prevent the FFP. Therefore, we concluded that it is not a field of the PDJ workflow design. It should be discussed as a matter of publication ethics.

The most significant loss for authors who receive a rejection notice is the time spent and the effort made to make the submission. The authors are expected to recover their costs by submitting their work to any other journal; however, this would require a similar review process. We expect the review processes in other journals to be simplified and to use processes, such as the cascading peer review mechanism [8].

*D. Data Repository*

The data repository is an information service for registering and publishing data. PDJ recommends using a trusted data repository that has a free access landing page, an open license policy, and a persistent identifier (e.g., a digital object identifier) publishing function. The data repository can directly manipulate the registered data through applications on the server. Therefore, the following data operations can occur:

- Data loss
- Data falsification
- Data fabrication

These are always present as information security risks for the data repository. These events would occur due to failure or unauthorized access to the information system. In this study, we thought that information security measures for the data repository are sufficient, and we want to discuss this possibility of occurrence from the unexpected activity related to each role in the review process.

If data is overwritten when it is updated, the past data may be lost. Data updates by authors (as shown in Section III-A) may result in the loss of the identity of the peer-reviewed data. End users can confirm the data integrity from the hash value on the peer review report. The most important thing here is whether the data before updating remains in the data repository. The core trustworthy data repository requirements include the version control strategy [9], but it does not describe the specific requirement. The required function is just that all pre-update versions are accessible. The download link of the reviewed data must be unique. The requirement of the data repository for the PDJ defined these functions. The PDJ does not allow the use of the private data repository or the Website.

Data falsification would occur due to improper operations or configuration on the data repository side. It often happens because of human error or setting error in the data repository side. Comparing the hash value at each stage under the review process (see Section III-A) detects data falsification. If there is a difference in the hash values, the PDJ secretariat immediately contacts the editor. The editor starts investigating data falsification. If an unauthorized access or attack is suspected, the PDJ secretariat immediately notifies the management organization of the data repository, and requests a response of the information incident.

Data fabrication can occur as a result of collusion between the data repository and the author. The occurrence of this event requires some mutually beneficial relationship between the data repository and the author. This act reduces the reliability of the data journal and creates a conflict of interest in the data repository. The independence of the repository also needs to be discussed, but this is not possible right now without sufficient information. Future research is expected about repository-driven frauds.

*E. Secretariat*

The secretariat is responsible for administrative procedures of the review process. The secretariat contacts each role, hash value calculation, data confirmation, and the creation of the peer review report. The secretariat operation proceeds according to the journal workflow. This role cannot operate the data directly. However, some losses may occur as a result of procedural errors or delayed actions; this is because the entire workflow is involved. PDJ has various services to automate the procedure to reduce human error.

*F. Other Services*

External information services are mostly used for a journal support system including submission and publishing. PDJ uses the EM for the review process and the JAIRO Cloud system for paper publishing. These services are not directly related to inappropriate data manipulation because these are independent of the PDJ editorial board. However, a general information security risk exists.

IV. CASE STUDY

PDJ released six articles in March 2019. Two inappropriate data manipulations have occurred since the launch of PDJ. Both manipulations were detected in the process of confirming the data identity using hash values before publication. We describe the survey results of two cases.

*A. Case 1: Download Link Generation Bug in Data Repository*

Despite downloading data from the data repository using the same link, different hash values were output each time they were downloaded. The secretariat notified the data repository administrator of this event. The administrator examined the download link on the landing page. As a result of the investigation, it was found that when a download is started on the landing page, compression processing of the registered data is started. Compressed data with different timestamps were created depending on the current date; therefore, the hash value of the downloaded data was altered. We concluded the case of the technical problem of download implementation of data repository. The secretariat requested the data repository to prepare a download link with the same hash value. The data repository was improved to suppress the data creation with different timestamps.

*B. Case 2: Data Not Present in Data Repository*

The author received comments from the referee and realized that it was necessary to correct the data. Therefore, the author directly contacted the data repository administrator to request withdrawing the old data and re-register the new data. The administrator did not recognize that the data was under review and processed it as requested by the author. The author responded only to the corrections of the manuscript at the time of revising and did not respond to the data correction.

When the manuscript was accepted, a comparison of the hash values revealed that the data had been modified after submission. The editorial chairperson asked the data repository administrator to clarify the details of the data update because of the possibility of serious data falsification. As a result, the above situation became clear.

After interviewing the author, we realized that the author was not aware of the prohibition on data modifications under the review process without notification. We concluded that this was not a malicious act because the author lacked the necessary information. The author was negligent without being malicious. The data repository performed high-level data manipulation including deletions and changes by directly contacting the author. For this reason, the PDJ secretariat requested the data repository to define strict operation rules and procedures. The secretariat decided always to compare the hash values of the data when revising the data.

## V. Summary

This study presents the possibility of inappropriate data manipulation in the PDJ. We also considered how to prevent this manipulation (Table I). Primary measures that we used, such as a strict policy and governance, were the same as the FFP measure for ordinary journals, but the data-specific part was given a new design. As shown in the case studies, a comparison of the hash values succeeded in detecting inappropriate data manipulation.

It is necessary to compare the hash values to detect unauthorized data changes during the review process. The secretariat calculates the hash value in the PDJ workflow, but the calculation flow is not integrated into the system as yet. It is implemented in the information services related to the data journal. In association with this requirement, the data repository also needs to have the hash value listed on the landing page and the data download link.

It is not yet clear how the peer review report would work because it aims at the social suppression effect for the referee and the editor. Further experimentation is required.

These measures will eventually become obsolete with advances in information technology. It is essential to maintain the reliability of data journals and data disclosure by accumulating various cases in the future.

ACKNOWLEDGMENT

We want to thank PDJ editorial board members and the secretariat for fully implementing these measures we designed for the data journal. Moreover, also we thank Dr. Akira Kadokura in Research Organization of Information System for feedback after the introduction.